\documentclass{article}
\usepackage{epsfig}
\newcommand{\be}{\begin{equation}}
 \newcommand{\ee}{\end{equation}}
 \newcommand{\bea}{\begin{eqnarray}}
 \newcommand{\eea}{\end{eqnarray}}
 \newcommand{\nn}{\nonumber}

\begin{document}

\begin{center}
{\Large \bf Polarization effects in the non-linear Compton
scattering\footnote{This work is supported in part by INTAS
and by the Fund of Russian Scientific Schools (code 2339.2003.2).}} \\

\vspace{4mm}

D.Yu.~Ivanov$^{1)}$, G.L.~Kotkin$^{2)}$,
V.G.~Serbo$^{2)}$\\
{\it $^{1)}$Sobolev Institute of Mathematics,
Novosibirsk, 6300090, Russia} \\
{\it $^{2)}$Novosibisk State University, Novosibirsk, 6300090,
Russia }

\end{center}

\begin{abstract}
 We consider emission of a photon by an electron in the
field of a strong laser wave. A probability of this process for
circularly or linearly polarized laser photons and for arbitrary
polarization of all other particles is calculated. We obtain the
complete set of functions which describe such a probability in a
compact invariant form. Besides, we discuss in some detail the
polarization effects in the kinematics relevant to the problem of
$e\to \gamma$ conversion at $\gamma \gamma$ and $\gamma e$
colliders.
\end{abstract}


\section{Introduction}

The analysis of polarization effects in the Compton scattering
 \be
  e(p) +\gamma (k) \to e(p') +\gamma (k')
  \label{1}
 \ee
is now included in text-books (see, for example, \cite{BLP}, \S
87). Nevertheless, the complete results for the cross sections
with both initial and final particles polarized has been obtained
only recently (see \cite{Grozin,GKPS83,KPS98} and the literature
therein). One interesting application of the process (\ref{1}) is
the collision of an ultra-relativistic electron with a beam of
polarized laser photons. In this case the Compton effect is the
basic process for obtaining of high-energy photons for
contemporary experiments in nuclear physics and for future $\gamma
\gamma $ and $\gamma e$ colliders \cite{GKST83}. The importance of
the particle polarization is clearly seen from the fact that in
comparison with the unpolarized case the number of final photons
with maximum energy is nearly doubled when the helicities of the
initial electron and photon are opposite~\cite{GKPS83}.

With the growth of the laser field intensity, an electron starts
to interact coherently with $n$ laser photons,
 \be
   e(q) +n\,\gamma_L (k) \to e(q') +\gamma (k')\,,
 \label{2}
 \ee
thus the Compton scattering becomes non-linear. Such a process
with absorption of $n=1,\, 2,\, 3,\, 4$ linearly polarized laser
photons was observed in the recent experiment at SLAC~\cite{SLAC}.
The polarization properties of the process (\ref{2}) are important
for future $\gamma \gamma $ and $\gamma e$ colliders
(see~\cite{GKPnQED,GKLT2001} and the literature therein). The
non-linear Compton scattering must be taken into account in
simulations of the processes in a conversion region of these
colliders. For comprehensive simulation, including processes of
multiple electron scattering, one has to know not only the
differential cross section of the non-linear Compton scattering
with a given number of the absorbed laser photons $n$, but energy,
angles and polarization of final photons and electrons as well.
The method of calculation for such cross sections was developed by
Nikishov and Ritus~\cite{NR-review}. It is based on the exact
solution of the Dirac equation in the field of the external
electromagnetic  plane wave. Some particular polarization
properties of this process were considered
in~\cite{NR-review}--\cite{BPM} and have already been included in
the existing simulation codes~\cite{Yokoya,TCode}.

In the paper~\cite{IKS04} we presented the complete description of
the non-linear Compton scattering for the case of circularly or
linearly polarized laser photons and arbitrary polarization of all
other particles. Besides, we derived (i) the approximate formulae
relevant for the problem of $e\to \gamma$ conversion; (ii) the
polarization of the final photons and electrons averaged over
their azimuthal angles; (iii) the limiting cases of the small and
large energies of the final photons; (iv) some numerical results
obtained for the range of parameters close to those in the
existing TESLA project~\cite{TESLA}. Here we present the short
review of the results obtained in \cite{IKS04}.

We use the system of units in which $c=1$, $\hbar=1$. In what
follows, we will often consider the non-linear Compton scattering
in the frame of reference in which a high-energy electron performs
a head-on collision with laser photons, i.e. in which ${\bf p}
\,\parallel \,(- {\bf k})$. We call this the ``collider system''.
(For the case of non-head-on collision of the initial particles
see ~\cite{IKS04}.)

\section{Kinematics}

Let us consider the interaction of an electron with a
monochromatic plane wave. The corresponding electric and magnetic
fields are ${\bf E}$ and ${\bf B}$, a frequency is $\omega$, and
let $F$ be the root-mean-squared field strength, $F^2 =\langle
{\bf B}^2 \rangle = \langle {\bf E}^2 \rangle$ and $n_{\rm L}$ be
the density of photons in the laser wave. The parameter describing
the intensity of the laser field (the parameter of non-linearity)
is defined as
 \be
\xi^2 = \left({e F \over m \omega }\right)^2={4\pi \alpha \over
m^2\,\omega}\, n_{\rm L}\,,
 \label{5}
  \ee
where $e$ and $m$ are the electron charge and the mass.

~From the classical point of view, the oscillated electron emits
harmonics with frequencies $n\,\omega$, where $n=1,\;2,\;\dots$
Their intensities at small $\xi^2$ are proportional to $({\bf
E}^2)^n \propto \xi^{2n}$, the polarization properties of these
harmonics depend on the polarizations of the laser wave and the
initial electron. From the quantum point of view, this radiation
can be described as the non-linear Compton scattering with
absorption of $n$ laser photons. When describing such a
scattering, one has to take into account that in a field of the
laser wave the 4-momenta $p$ and $p'$ of the free initial and
final electrons are replaced by the 4-quasi-momenta $q$ and $q'$,
 \be
q=p+\xi^2{ m^2\over 2pk}\,k\,,\;\; q'=p'+\xi^2 {m^2\over
2p'k}\,k\,,\;\;
 q^2=(q')^2= (1+\xi^2 )\,m^2\,.
  \label{6}
 \ee
As a result, we deal with the reaction (\ref{2}) for which the
conservation law reads
 \be
 q+n\,k=q'+k'\,.
 \label{8}
 \ee
Since $qk=pk$, it is convenient to use the same invariant
variables  as for the linear Compton scattering (compare
\cite{GKPS83}):
 \be
x={2p k\over m^2}\approx {4E\omega\over m^2} \,, \;\;\; y\; = {k
k'\over p k}\approx {\omega' \over E} \leq y_n = {nx\over
1+nx+\xi^2}\,.
 \label{9}
 \ee

The invariant description of the polarization properties of both
the initial and the final photons can be performed in the standard
way (see \cite{BLP}, \S 87). We define a pair of unit 4-vectors
 \bea
e^{(1)}&=&{N\over\sqrt{-N^2}}\,,\;\;\;
e^{(2)}={P\over\sqrt{-P^2}}\,,
  \label{19}
  \\
N^\mu&=&\varepsilon^{\mu\alpha \beta \gamma} P_\alpha
(k'-n\,k)_\beta K_\gamma \, ,\;\;
  P_\alpha=(q+q')_\alpha-{(q+q')K\over
K^2}\,K_\alpha \,,\;\;\; K_\alpha=n\,k_\alpha+k'_\alpha\,.
 \nn
 \eea
The 4-vectors $e^{(1)}$ and $e^{(2)}$ are orthogonal to each other
and to the 4-vectors $k$ and $k'$, therefore, they are fixed with
respect to the scattering plane of the process.

Let $\xi_j$ be the Stokes parameters for the initial photon which
are defined with respect to 4-vectors $e^{(1)}$ and $e^{(2)}$. As
for the polarization of the final photon, it is necessary to
distinguish the polarization $\xi_j^{(f)}$ of the final photon as
resulting from the scattering process itself from the detected
polarization $\xi '_j$ which enters the effective cross section
and which essentially represents the properties of the detector as
selecting one or other polarization of the final photon (for
detail  see~\cite{BLP}, \S 65). Both these Stokes parameters,
$\xi_j^{(f)}$ and $\xi '_j$, are also defined with respect to the
4-vectors $e^{(1)}$ and $e^{(2)}$.

Let {\boldmath $\zeta$} be the polarization vector of the initial
electrons. As with the final photon, it is necessary to
distinguish the polarization {\boldmath $\zeta$}$^{(f)}$ of the
final electron as such from the polarization {\boldmath $\zeta
$}$^{\prime}$ that is selected by the detector. The vectors
{\boldmath $\zeta$} and {\boldmath $\zeta $}$^{\prime}$ enter the
effective cross section. They also determine the electron-spin
4-vectors $a$ and $a^{\prime}$.

Now we have to define invariants which describe the polarization
properties of the initial and the final electrons. For the linear
Compton scattering, the relatively simple description was obtained
in ~\cite{Grozin} using invariants which have a simple meaning in
the center-of-mass system. However, this frame of reference is not
convenient for the description of the non-linear Compton
scattering, since it has actually to vary with the change of the
number of the absorbed laser photons $n$.

Since $ap=a'p'=0$, we can decompose the electron polarization
4-vectors over three convenient unit 4-vectors $e_j$ and $e_j'$,
$j=1,\,2,\,3$, projections on which determine polarization
properties of the electrons. Our choice is based on the experience
obtained in~\cite{KPS98} and \cite{BPM}:
 \bea
e_1&=&e^{(1)}\,,\;\;\; e_2=-e^{(2)}-{\sqrt{-P^2}\over m^2
x}k\,,\;\;\;\;\;\;\;\;\; e_3={1\over m}\left(p-{2\over x}k\right);
 \label{23}\\
e^{\prime}_1&=&e^{(1)}\,,\;\;\;e^{\prime}_2=-e^{(2)}-{\sqrt{-P^2}
\over m^2 x(1-y)}k\,,\;\; e^{\prime}_3={1\over
m}\left(p^{\prime}-{2\over x(1-y)}k\right).
 \nn
 \eea
These vectors satisfy the conditions
 \begin{equation}
e_i e_j=-\delta_{ij}\;,\;\;\;e_j p=0\; ; \;\;\;e^{\prime}_i
e^{\prime}_j= -\delta_{ij}\; , \;\;\; e^{\prime}_j p^{\prime}=0\,.
 \label{24}
\end{equation}
It allows us to represent the 4-vectors $a$ and $a^{\prime}$ in
the following covariant form: $a=\sum_{j=1}^3 \zeta_j e_j$,\\
$a^{\prime}=\sum_{j=1}^3 \zeta^{\prime}_j e^{\prime}_j$, where
 \be
 \zeta_j=-ae_j\,,\;\;\;
\zeta^{\prime}_j=-a^{\prime}e^{\prime}_j\,.
 \label{26}
\end{equation}

\section{Cross section in the invariant form}

The usual notion of the cross section is not applicable for the
reaction (\ref{2}) and usually its description is given in terms
of the probability of the process per second $\dot{W}^{(n)}$.
However, for the procedure of simulation in the conversion region
as well as for the simple comparison with the linear case, it is
useful to introduce the ``effective cross section'' given by the
definition
 \be
d\sigma^{(n)}= {d\dot{W}^{(n)}\over j}\,,\;\; j= {(q\,k) \over
q_0\, \omega}\,n_{\rm L} ={m^2 \,x\over 2q_0 \omega} \, n_{\rm
L}\,,
 \label{17}
 \ee
where $j$ is the flux density of colliding particles. Contrary to
the usual cross section, this effective cross section does depend
on the laser beam intensity, i.e. on the parameter  $\xi^2$. The
total effective cross section is defined as the sum over
harmonics, corresponding to the reaction (\ref{2}) with a given
number $n$ of the absorbed laser photons:
  \be
d\sigma=\sum\limits_{n}d\sigma^{(n)}\,.
 \label{18}
 \ee

The effective differential cross section can be presented in the
following invariant form:
 \be
d\sigma (\mbox{\boldmath$\zeta$},\,\mbox{\boldmath$\xi$},\,
\mbox{\boldmath$\zeta$}^{\prime},\,\mbox{\boldmath$\xi$}^{\prime})
= {r_e^2\over 4x}\;\sum_n F^{(n)}\;d\Gamma_n\,,\;\;d\Gamma_n =
\delta (q+n\,k-q'-k')\;{d^3k'\over \omega'}{d^3q'\over q_0'}\,,
  \label{33}
 \ee
where $r_e=\alpha/m$ is the classical electron radius, and
 \be
F^{(n)}=F_0^{(n)}+\sum ^3_{j=1}\left( F_j^{(n)}\xi '_j\; +
\;G_j^{(n)} \zeta^{\prime}_j\right)
 + \sum ^3_{i,j=1}H_{ij}^{(n)}\,\zeta^{\prime}_i\,\xi^{\prime}_j
\,.
 \label{34}
 \ee
In the collider system
 \be
d\Gamma_n = dy\, d\varphi\,,
 \ee
where $\varphi$ is the azimuthal angle of the final photon. The
function $F_0^{(n)}$ describes the total cross section for a given
harmonic $n$, summed over spin states of the final particles:
 \be
\sigma^{(n)}(\mbox{\boldmath$\zeta$},\,\mbox{\boldmath$\xi$})=
{r_e^2\over x}\; \int F_0^{(n)}\,d\Gamma_n \,.
 \label{35}
 \ee
The terms $F_j^{(n)}\xi '_j$  and $G_j^{(n)} \zeta^{\prime}_j$ in
(\ref{34}) describe the polarization of the final photons and the
final electrons, respectively. The last terms $H_{ij}^{(n)}
\zeta^{\prime}_i\,\xi^{\prime}_j$ stand for the correlation of the
final particles' polarizations.

From (\ref{33}), (\ref{34}) one can deduce the polarization of the
final photon $\xi_j^{(f)}$ and electron $\zeta^{(f)}_j$ resulting
from the scattering process itself. According to the usual rules
(see~\cite{BLP}, \S 65), we obtain the following expression for
the Stokes parameters of the final photon (summed over
polarization states of the final electron):
 \be
\xi_j^{(f)}= {F_j\over F_0}\,,\;\; F_0=\sum_n
F_0^{(n)}\,,\;\;F_j=\sum_n F_j^{(n)}\,;\;\;  j= 1,\,2,\,3\,.
 \label{36}
 \ee
The polarization of the final electron (summed over polarization
states of the final photon) is given by invariants
 \be
\zeta_j^{(f)}= {G_j\over F_0}\,,\;\;G_j=\sum_n G_j^{(n)}\,.
 \label{37}
 \ee
In the similar way, the polarization properties for a given
harmonic $n$ are described by
\begin{equation}
\xi_{j}^{(n)(f)}= {F_j^{(n)}\over F_0^{(n)}}\,,\;\;
\zeta_{j}^{(n)(f)}= {G_j^{(n)}\over F_0^{(n)}}\,.
 \label{39}
\end{equation}

As an example, we present here the functions $F_{0,1,2,3}$ for the
circularly polarized laser photons. In this case the
electromagnetic laser field is described by the 4-potential
 \be
A_\mu(x) ={m\over e}\,\xi \;\left[e_\mu^{(1)}\,
\cos{(kx)}+P_c\,e_\mu^{(2)}\, \sin{(kx)} \right]\,,
 \label{2.2a}
 \ee
where the unit vectors $e_\mu^{(1,2)}$ are given in (\ref{19}) and
$P_c$ is the degree of the circular polarization of the laser wave
or the initial photon helicity. We have calculated the
coefficients $F_j^{(n)},\, G_j^{(n)}$ and $H_{ij}^{(n)}$ using the
standard technique presented in~\cite{BLP}, \S 101. The necessary
traces have been calculated using the package MATHE\-MA\-TICA. In
the considered case of the 100 \% circularly polarized ($P_c=\pm
1$) laser beam, almost all dependence on the non-linearity
parameter $\xi^2$ accumulates in three functions:
  \bea
f_n& \equiv&
f_n(z_n)=J_{n-1}^{2}(z_n)+J_{n+1}^{2}(z_n)-2J_{n}^{2}(z_n)\,,
 \nn\\
 g_n& \equiv& g_n(z_n)=\frac{4 n^2 J_{n}^2(z_n)}{z_n^2}\,,
 \label{40}\\
h_n&\equiv& h_n(z_n)=J_{n-1}^{2}(z_n)-J_{n+1}^{2}(z_n)\,,
 \nn
 \eea
where $J_n(z)$ is the Bessel function. The functions (\ref{40})
depend on $x$, $y$ and $\xi^2$ via the single argument
 \be
z_n= {\xi\over \sqrt{1+\xi^2}}\; n\,s_n\,,
 \label{41}
 \ee
where
 \begin{equation}
s_n=2\sqrt{r_n(1-r_n)},\;\; c_n= 1-2r_n\,,\;\;
r_n={y\,(1+\xi^2)\over (1-y)\,nx}\,.
  \label{10}
\end{equation}

For the small value of this argument one has \be
f_n=g_n=h_n=\frac{(z_n/2)^{2(n-1)}}{[(n-1)!]^2} \;\; \mbox{ at}
\;\; z_n \to 0\,,
 \label{42}
 \ee
in particular,
 \be
 f_1=g_1=h_1=1
\;\; \mbox{ at} \;\; z_1= 0\,.
 \label{43}
 \ee

The results of our calculations are the following. The function
$F_0^{(n)}$, related to the total cross section (\ref{35}), reads
 \be
F_0^{(n)}=\left({1\over 1-y}+1-y\right)\,f_n- {s_n^2\over
1+\xi^2}\, g_n- \left[{y s_n \over \sqrt{1+\xi^2}}\, \zeta_2
-{y(2-y)\over 1-y}\, c_n \,\zeta_3\right] \,h_n\,P_c\,.
 \label{45}
 \ee
The polarization of the final photons $\xi_j^{(f)}$ is given by
Eq. (\ref{36}) where
  \bea
F_1^{(n)}&=&{y\over 1-y} {s_n \over \sqrt{1+\xi^2}}\, h_n P_c\,
\zeta_1\,,
 \label{46}
 \\
F_2^{(n)}&=&\left({1\over 1-y} +1-y\right)\, c_n h_n P_c- {ys_n
c_n \over \sqrt{1+\xi^2}}\, g_n \, \zeta_2+y\left( {2-y\over
1-y}\,f_n- {s^2_n\over 1+\xi^2}\,g_n\right) \zeta_3\,,
 \nn
 \\
F_3^{(n)}&=&2(f_n-g_n)+ s^2_n (1+\Delta)\,g_n-{y\over 1-y} {s_n
\over \sqrt{1+\xi^2}}\, h_n P_c\, \zeta_2
 \nn
 \eea
and $\Delta= \xi^2/ (1+\xi^2)$. Note that our results for
$F_0^{(n)}$ and $F_2^{(n)}$ coincide with that ones in the
literature, whereas the results for $F_1^{(n)}$ and $F_3^{(n)}$
are new.

\section{Examples and discussion}

In  this section we give some examples which illustrate the
dependence of the differential cross sections and the
polarizations on the final photon energy. We restrict ourselves to
properties of the high-energy photon beam which are of most
importance for the future $\gamma \gamma$ and $\gamma e$
colliders. It is expected (see, for example, the TESLA
project~\cite{TESLA}) that in the conversion region of these
colliders, an electron with the energy $E= 250$ GeV performs a
head--on collisions with laser photons having the energy $\omega
\approx 1$ eV per a single photon. The parameters used below are
close to those in the TESLA project. In particular,
 $$
 x\approx 4E\omega/m^2=4.8\,,
 $$
and the non-linearity parameter $\xi^2$ (\ref{5}) is chosen either
the same as in the TESLA project, $\xi^2=0.3$, or larger by one
order of magnitude, $\xi^2=3$, just to illustrate the tendencies.
In figures below we use notation
 $$
\sigma_0=\pi r_e^2 \approx 2.5\cdot 10^{-25} \;\;\mbox{cm}^2\,.
 $$

\begin{figure}[!htb]
\includegraphics[width=0.47\textwidth]{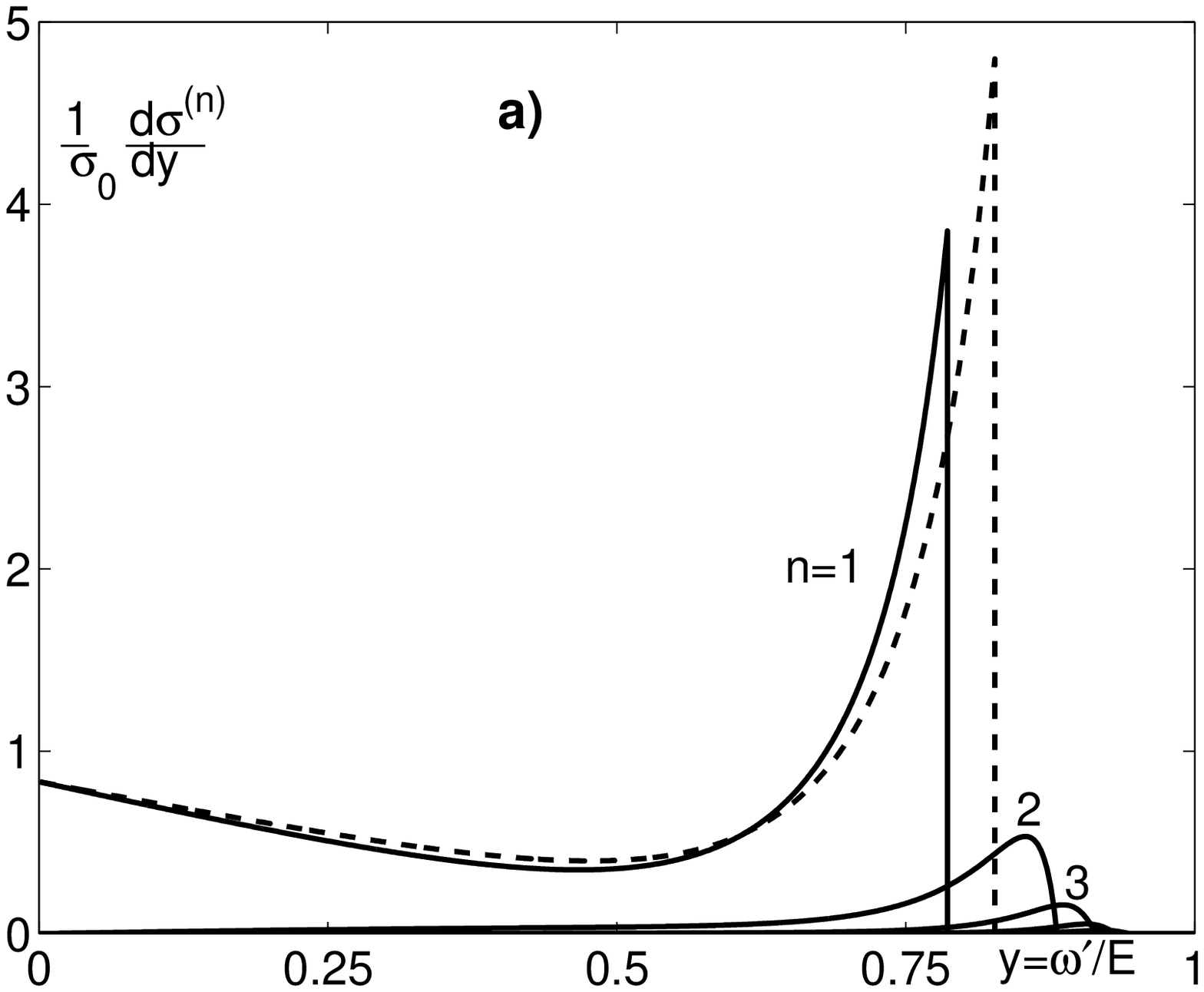}
\includegraphics[width=0.47\textwidth]{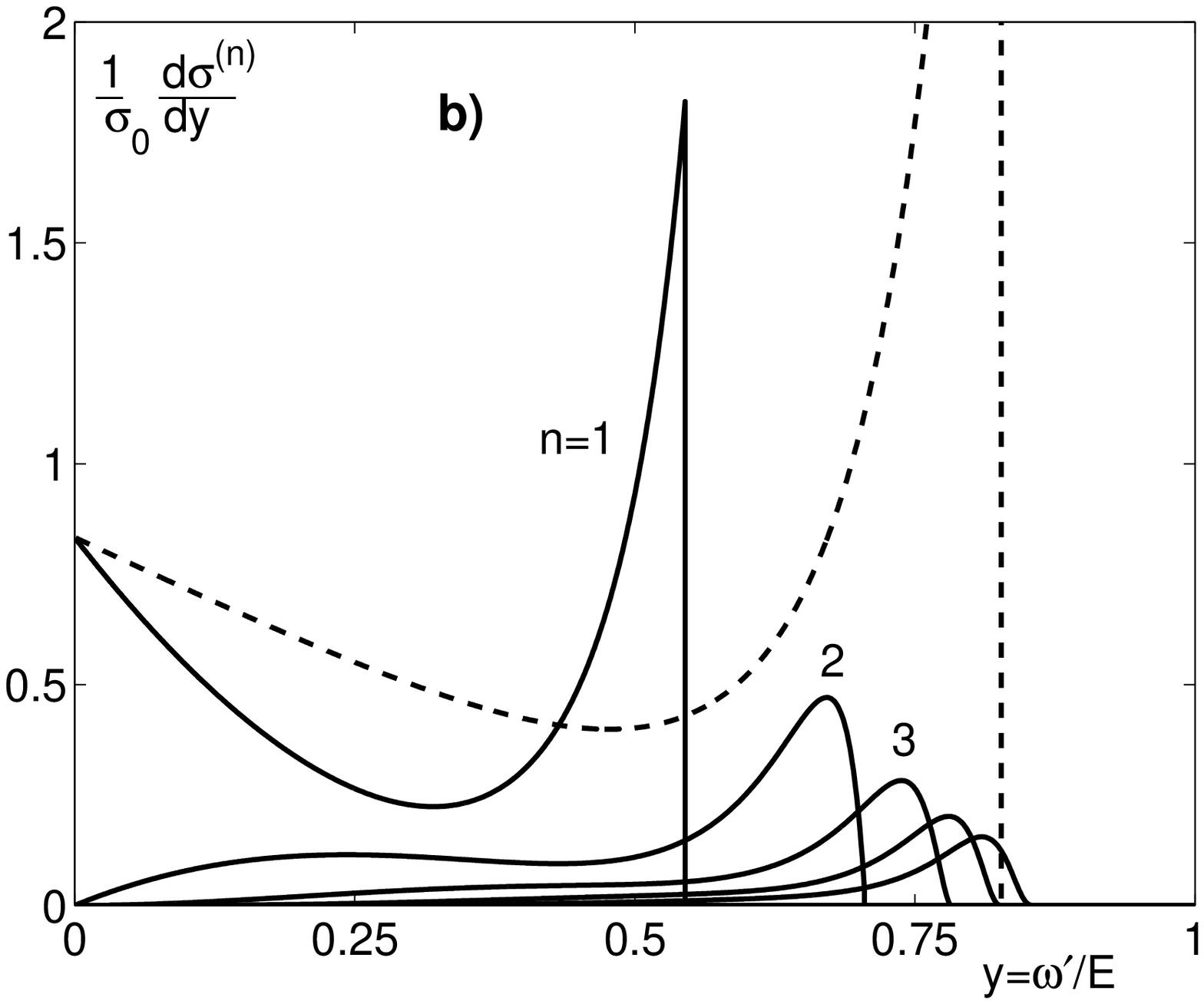}
 \caption{Energy spectra of final photons for different
harmonics $n$ at $\zeta_3\,P_c =-1$ and: {\bf (a)} $\xi^2=0.3$;
$\;$ {\bf (b)} $\xi^2=3$. The dashed curves correspond to
$\xi^2=0$. }
 \label{f1}
\end{figure}

{\it The case of the circularly polarized laser photons (Fig.
\ref{f1}).}

The spectra of the few first harmonics are shown in Fig. \ref{f1}
for the case of ``a good polarization'', when helicities of the
laser photon and the initial electron are opposite,
$\zeta_3P_c=-1$. At a small intensity of the laser wave ($\xi^2 =
0.3$, Fig. 1a) the main contribution is given by photons of the
first harmonic and the probability for generation of the higher
harmonics is small. However, with the growth of the non-linearity
parameter ($\xi^2 = 3$, Fig. 1b), the maximum energy for the first
harmonic decreases and the peak of this harmonic at $y=y_1$
decreases as well. As for the higher harmonics, with the rise of
$\xi^2$ (Fig. 1b) we see an increase of the yield of photons with
energies higher than the maximum energy of the first harmonic. As
a result, the total spectrum becomes considerable wider.

\begin{figure}[!htb]
\includegraphics[width=8cm]{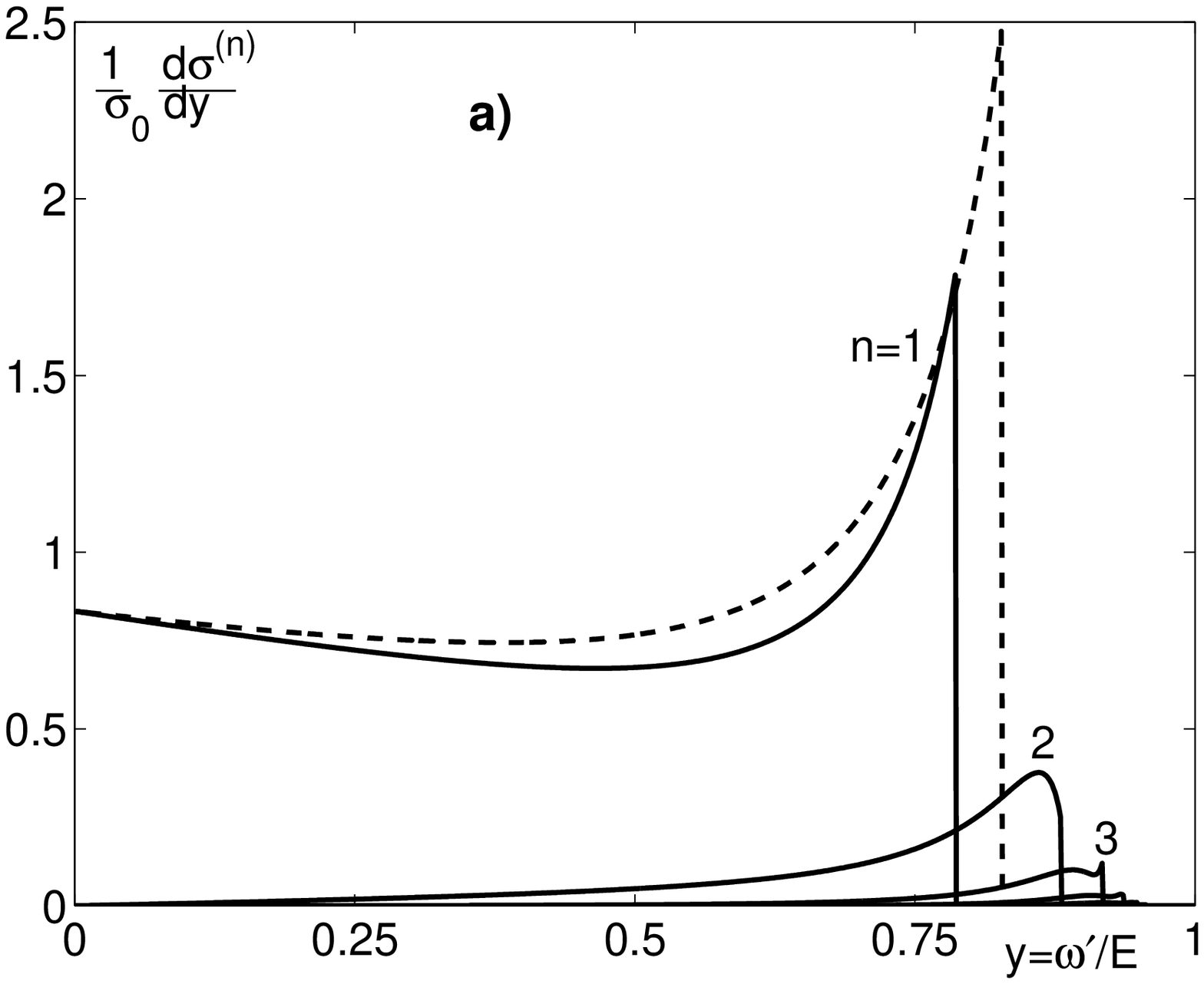}
\includegraphics[width=8cm]{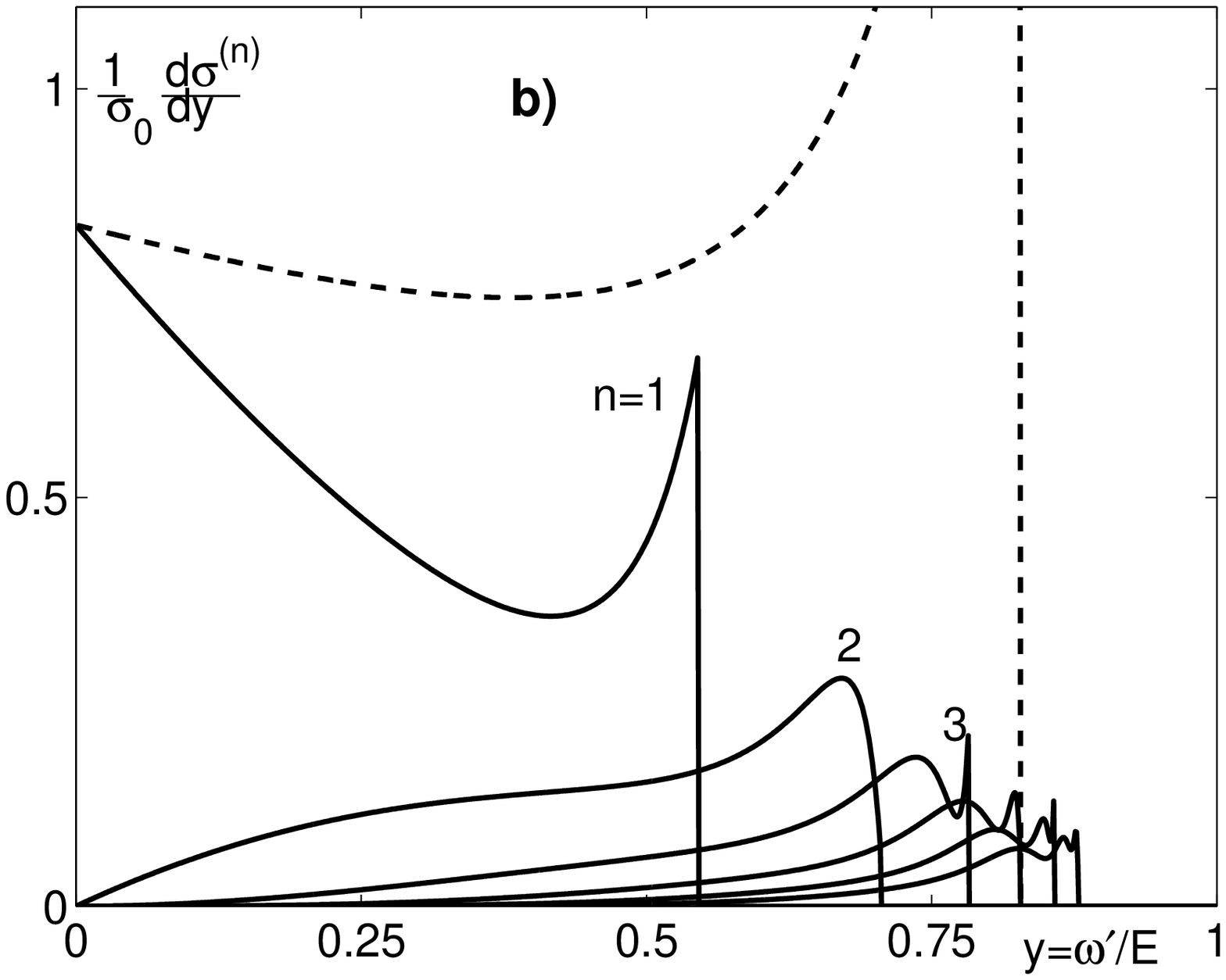}
\caption{Energy spectra of final photons for different harmonics
$n$ in the case of the linearly polarized laser photons and: {\bf
(a)} $\xi^2=0.3$; $\;$ {\bf (b)} $\xi^2=3$. The dashed curves
correspond to $\xi^2=0$.}
 \label{f2}
\end{figure}

\begin{figure}[!htb]
\includegraphics[width=0.47\textwidth]{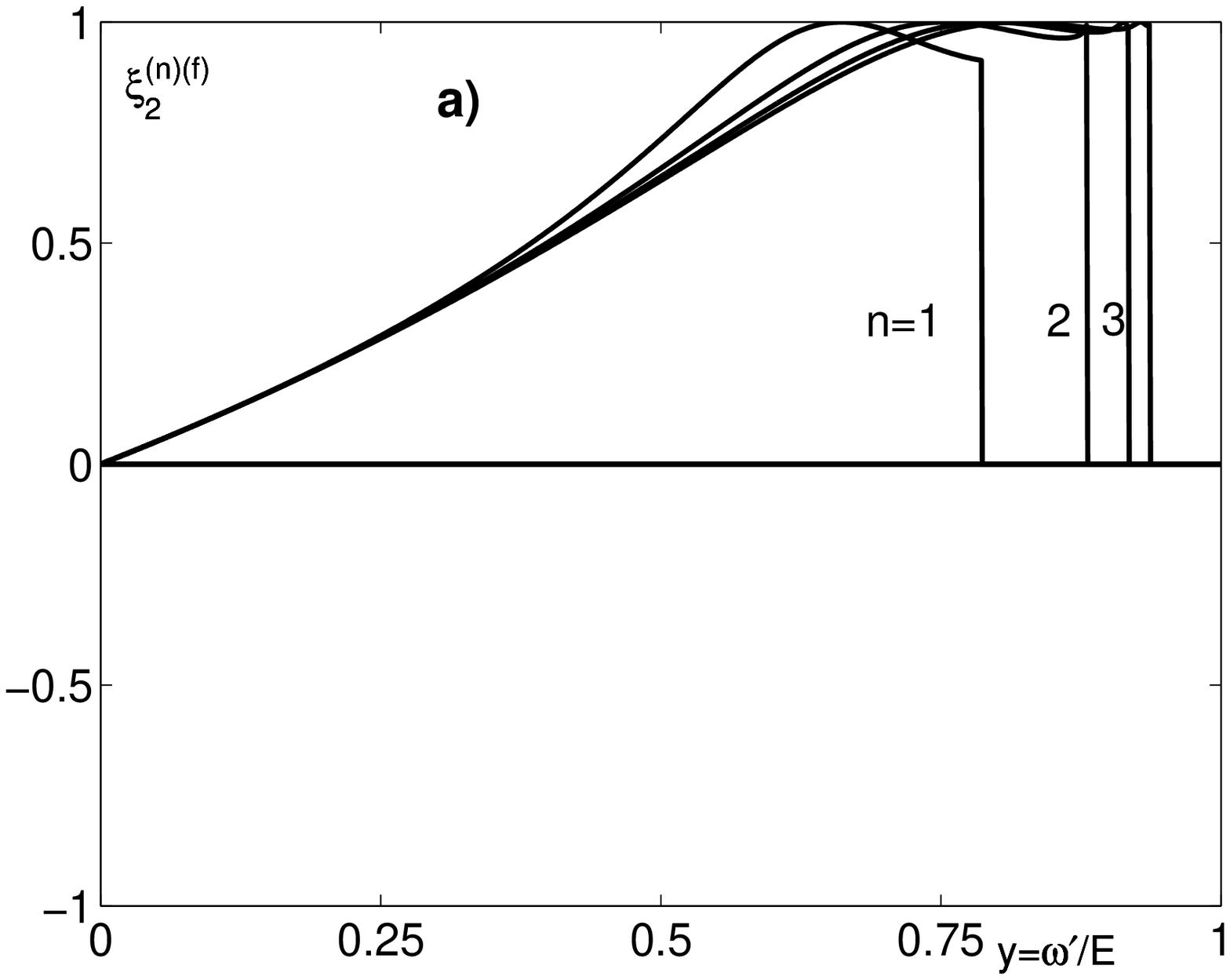}
\includegraphics[width=0.47\textwidth]{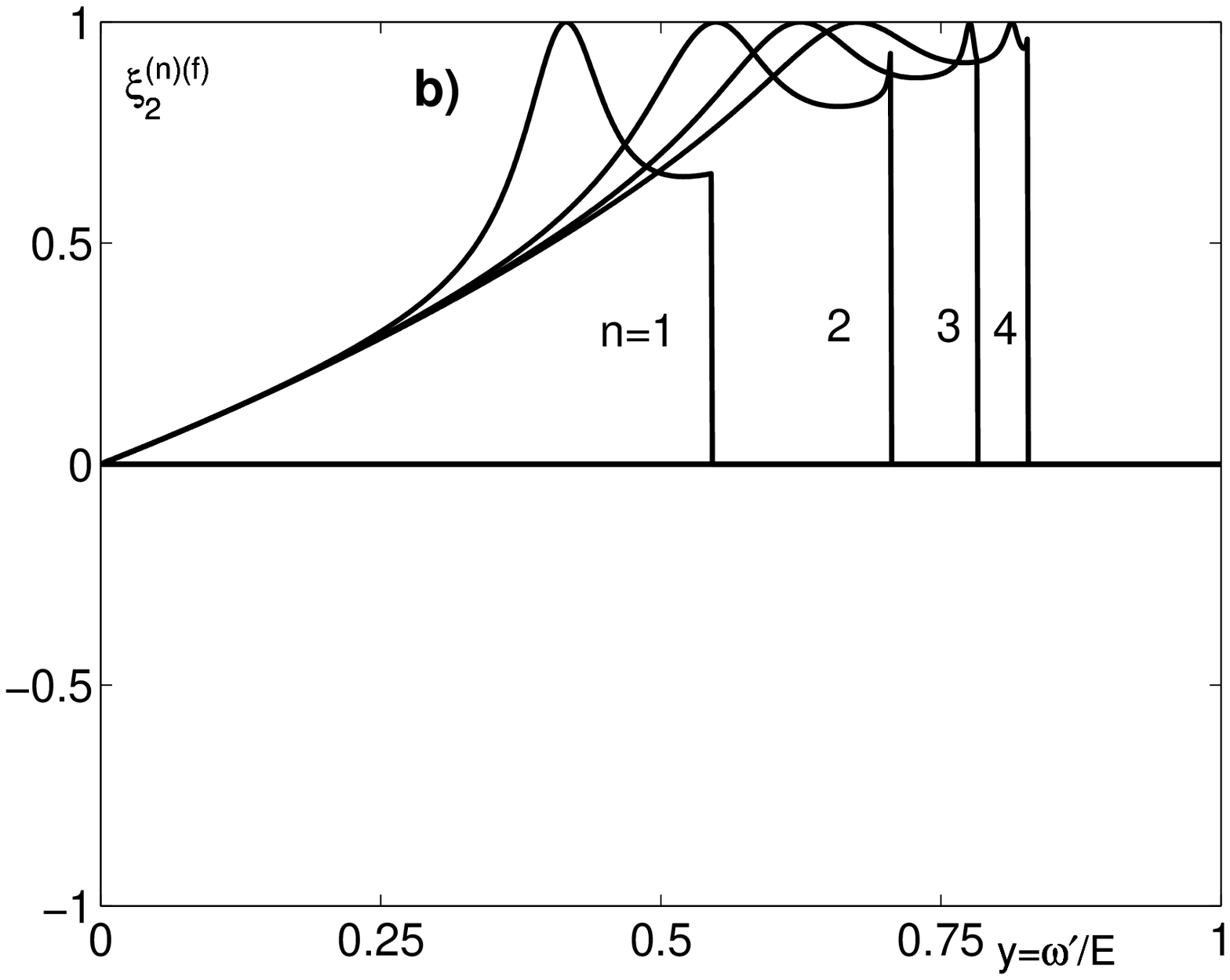}
\caption{The mean helicity $\xi_2^{(n)(f)}$ of the final photons
for different harmonics $n$ versus the final photon energy
$\omega'$ at $\zeta_3=1$ and: {\bf (a)} $\xi^2=0.3$; $\;$ {\bf
(b)} $\xi^2=3$. The laser photons are linearly polarized. The
scattering plane is parallel to the direction of the laser photon
polarization.}
 \label{f3}
\end{figure}

\begin{figure}[!htb]
\includegraphics[width=0.47\textwidth]{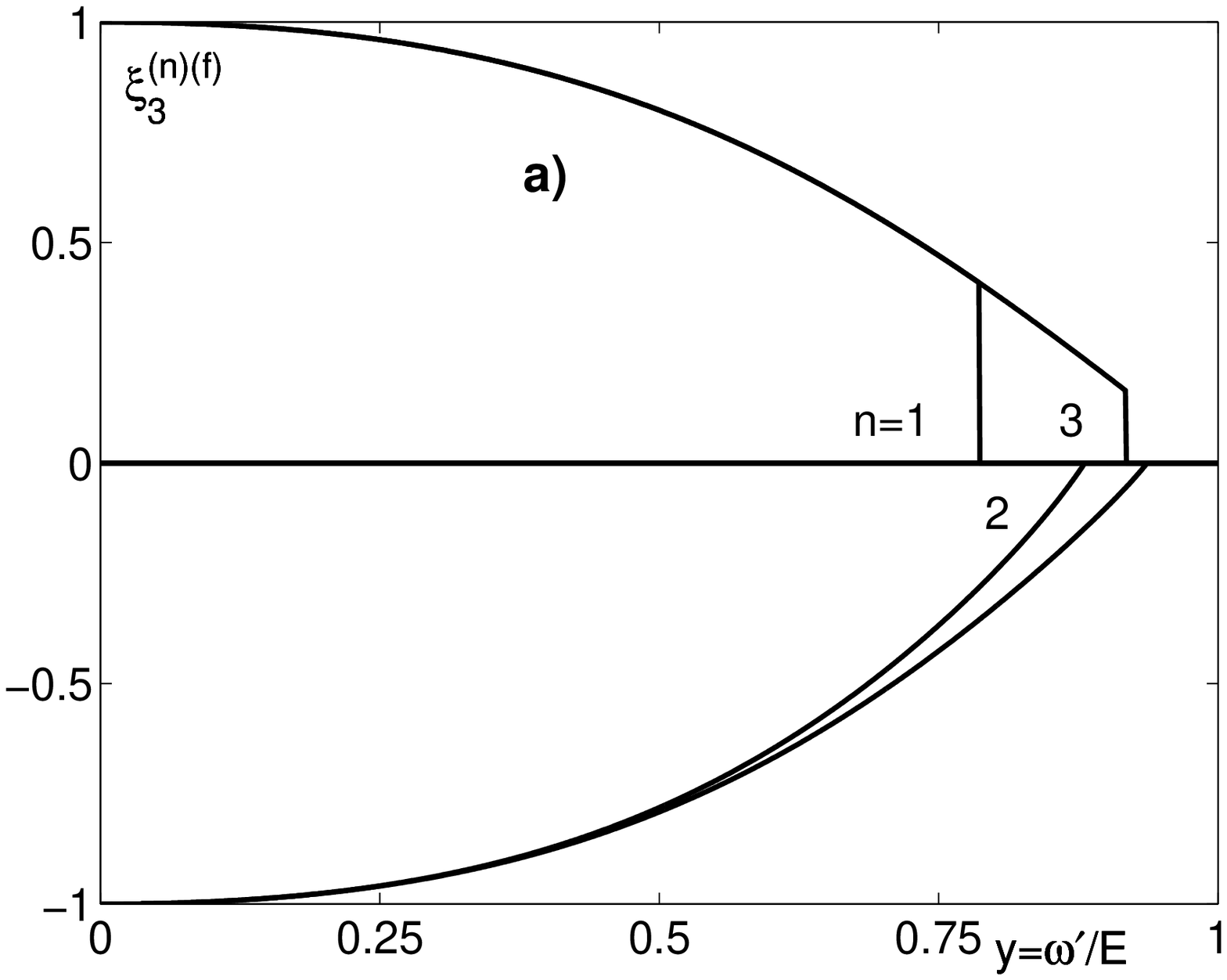}
\includegraphics[width=0.47\textwidth]{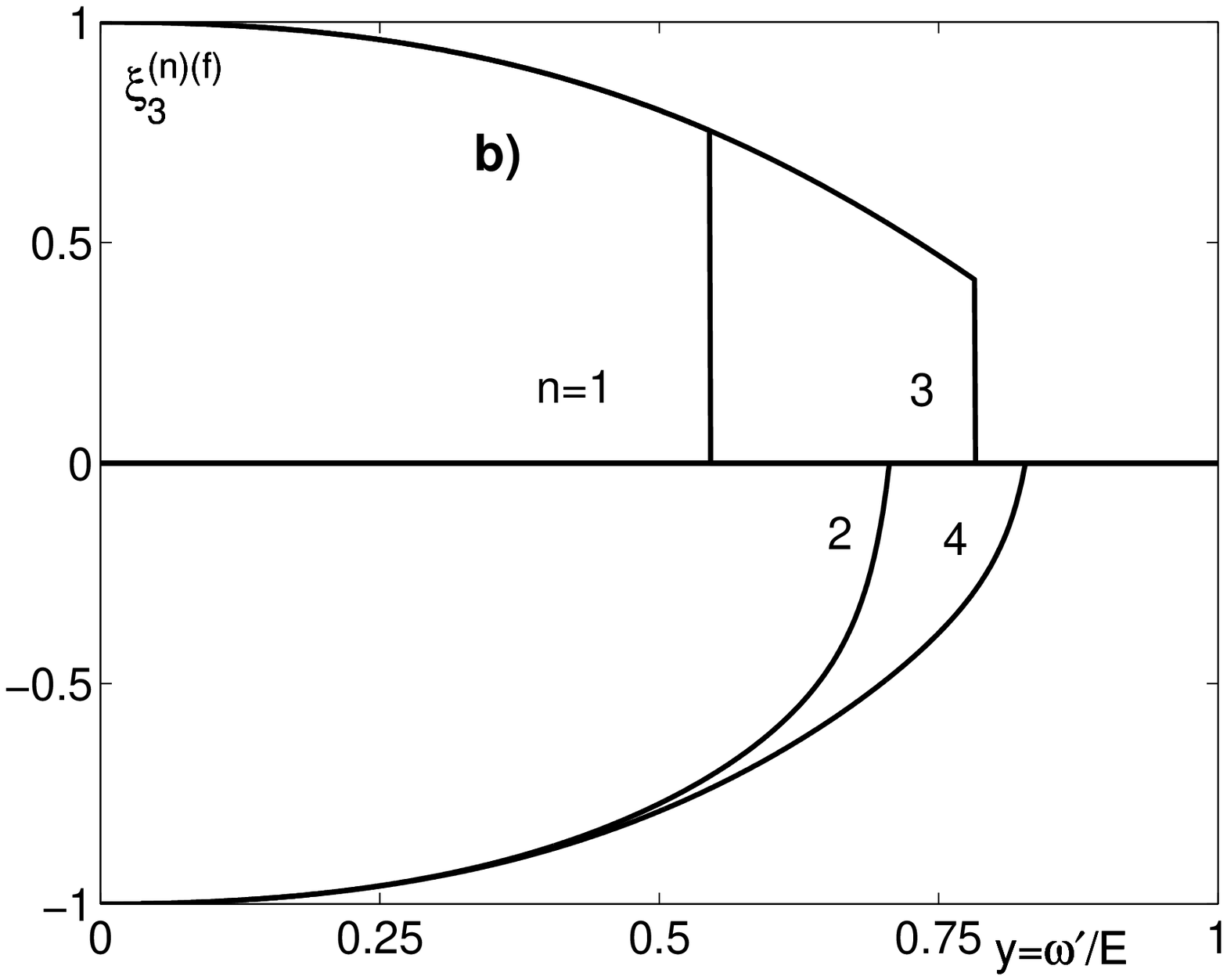}
\caption{The Stokes parameters $\xi_3^{(n)(f)}$ of the final
photons for different harmonics $n$ versus the final photon energy
$\omega'$ at $\zeta_3=1$ and: {\bf (a)} $\xi^2=0.3$; $\;$ {\bf
(b)} $\xi^2=3$. The laser photons are linearly polarized. The
scattering plane is perpendicular to the direction of the laser
photon polarization. The Stokes parameter $\xi_1^{(n)(f)}=0$ in
this case. }
 \label{f4}
\end{figure}

{\it The case of the linearly polarized laser photons (Figs.
\ref{f2}, \ref{f3}, \ref{f4}).}

The spectra of the first few harmonics for this case  are shown in
Fig. \ref{f2}. They differ considerably from those for the case of
the circularly polarized laser photons shown in Fig. 1. First of
all, in the considered case the spectra do not depend on the
polarization of the initial electrons. The maximum of the spectrum
for the first harmonic at $y=y_1$ now is about two times smaller
than that on Fig. 1. Besides, the harmonics with $n>1$ do not
vanish at $y=y_n$ contrary to such harmonics on Fig. 1.

For the first few harmonics the mean helicities of the final
photons, $\xi_2^{(n)(f)}$, are shown at $\zeta_3=1$  in Fig. 3a
for $\xi^2=0.3$ and in Fig. 3b for $\xi^2=3$. We would like to
direct attention to the surprising fact that each harmonic is
almost 100\% circularly polarized near the high-energy part of the
spectrum. The curves on Fig. \ref{f3} are given for the azimuthal
angle $\varphi=0$, for other values of $\varphi$ these curves
would look a bit different, but not too much.

The degree of linear polarization of the final photons is not
large in the high-energy part of the spectrum, but it becomes
rather high in the middle and the low part of the spectrum; see
Fig. \ref{f4}. Certainly, the direction of this polarization
depends on the azimuthal angle $\varphi$, as the result the linear
polarization averaged over $\varphi$ is substantially smaller than
one on Fig. 4. Nevertheless, the non-trivial effects, related to
the high degree of linear polarization present at a certain
$\varphi$, do exist. For example, it was shown \cite{PST2003} that
for linear Compton scattering this feature leads to an important
effect for the luminosity of $\gamma\gamma$ collisions.


%


\begin{thebibliography}{99}
\bibitem{Yokoya}
K.~Yokoya, CAIN2e Users Manual. The latest version is available
from ftp://lcdev.kek.jp/pub/Yokoya/cain235/CainMan235.pdf.
 \vspace{-2.5mm}
\end{thebibliography}
\end{document}